\documentclass[10pt,twocolumn,superscriptaddress,aps,prl,preprintnumbers,amsmath,amssymb,floatfix]{revtex4}

\usepackage[latin9]{inputenc}
\usepackage{color}

\usepackage{braket}
\usepackage{graphicx}
\usepackage{esint}
\usepackage[unicode=true,  bookmarks=false,  breaklinks=false,pdfborder={0 0 1},backref=false,colorlinks=true]  {hyperref}

\makeatletter
\@ifundefined{textcolor}{}
{%
 \definecolor{BLACK}{gray}{0}
 \definecolor{WHITE}{gray}{1}
 \definecolor{RED}{rgb}{1,0,0}
 \definecolor{GREEN}{rgb}{0,1,0}
 \definecolor{BLUE}{rgb}{0,0,1}
 \definecolor{CYAN}{cmyk}{1,0,0,0}
 \definecolor{MAGENTA}{cmyk}{0,1,0,0}
 \definecolor{YELLOW}{cmyk}{0,0,1,0}
}

\usepackage{graphicx}
\usepackage[english]{babel}
\usepackage{amsmath}
\usepackage{amssymb}
\definecolor{orange}{rgb}{1,0.5,0}
\usepackage{bm}

\newcommand{\bx}{{\mathbf x}}
\newcommand{\bk}{{\mathbf k}}
\newcommand{\bphi}{\bm \phi}
\newcommand{\bpi}{\bm \Pi}
\newcommand{\bG}{\bm G}

\begin{document}
\title{Quantum Quench of the ``Speed of Light'': \\ Quantum Dynamical Universality Classes and Short-time Universal Behavior}
\author{Mohammad F. Maghrebi}
\affiliation{Department of Physics and Astronomy, Michigan State University, East Lansing, Michigan 48824, USA}

\begin{abstract}
A long-lived prethermal state may emerge upon a sudden quench of a quantum system.
In this paper, we study a quantum quench of an initial {\it critical} state, and show that the resulting prethermal state exhibits a genuinely quantum and dynamical universal behavior. Specifically, we consider a scenario where the ``speed of light'' characterizing the propagation of local perturbations is suddenly quenched at criticality.
We also find that the system approaches the prethermal state in a universal way described by a new exponent that characterizes a kind of quantum aging.
\end{abstract}

\maketitle

The dynamics of quantum many-body systems following a sudden change of the Hamiltonian, also called a {\it quantum quench},
is one of the most active areas of nonequilibrium quantum physics \cite{polkovnikov11}.
It is widely believed that, after a quantum quench, isolated systems generically approach a thermal state (many-body localized systems being an exception) \cite{Deutsch91,Srednicki94,Rigol08,polkovnikov11, Rigol16review, Vasseur16}.
However, a long-lived quasi-stationary {\it prethermal} state may emerge before the onset of thermalization \cite{Berges04}; this typically occurs when there are  approximate conserved quantities preventing a complete loss of memory of the initial state and an immediate thermalization \cite{Kollar11,Kehrein08,Kehrein10,Marino12,Mitra13,Worm13,Marcuzzi13}. Prethermalization has also been observed in experiments \cite{Gring12,langen13,Langen15}.

Just like their thermal counterparts, prethermal states can exhibit universal scaling relations \cite{Schmiedmayer16review,Nowak12,Berges12,Karl13,Lux14,Berges08,Scheppach10}.
In analogy with thermal phases where a critical state can be reached by tuning the temperature to its critical value, a {\it prethermal critical state} emerges by tuning the control parameter of the quench.
These critical states exhibit long-range correlations, within a long time window,
with distinct universal properties \cite{Eckstein09,Hackl09,Pletyukhov10,Sciolla10,Gambassi11,Schiro10,Schiro11,Sciolla11,Sciolla13,Chandran13,Smacchia15,Mitra-Gambassi15,Gambassi-Mitra15,Mitra-Gambassi16,Chichinadze16,Maraga16,Marino17Prethermal}.
Moreover, the approach to the (prethermal) critical point leads to a distinct short-time universal behavior as the system still retains its memory of the initial state, and exhibits a kind of {\it quantum aging}; the analog of this phenomenon is also studied in open quantum systems \cite{Schmalian14,Schmalian15,Buchhold15,Lang16}.
Despite recent progress,
it appears that quantum features are at least partially obscured upon a quench in a closed system due to the emergence of an effective temperature \cite{Gambassi-Mitra15,Mitra-Gambassi15,Mitra-Gambassi16}. As a result, short-time universal behavior exhibits features that resemble those of {\it classical} aging \cite{Janssen89,Gambassi05review}.

In this work, we propose a quench protocol that goes beyond the paradigm of partial or full thermalization, and gives rise to a genuinely quantum dynamical universality class and a new type of short-time universal behavior. Specifically, we consider a many-body quantum system where the speed of light characterizing the propagation of local perturbations is suddenly quenched at criticality, and reveal the rich, universal dynamics before the onset of thermalization.

We first review a simple argument for why an effective temperature {\it generically} arises following a sudden quench, and discuss later how such a scenario can be evaded.
A paradigmatic example is a system of weakly interacting (bosonic) particles. Let us ignore interactions, and take the quasiparticle spectrum as $\omega_{0\bk}$ and $\omega_\bk$ before and after the quench, respectively.
As different modes are decoupled in a non-interacting system, we simply need to consider a collection of harmonic oscillators upon a sudden change of their natural frequency $\omega_{0\bk}\to \omega_\bk$ in the Hamiltonian $H=\sum_{\bk}|\Pi_\bk|^2 + \omega_\bk^2|\phi_\bk|^2$ where $\phi_\bk$ and $\Pi_\bk$ represent the (bosonic) field and its conjugate momentum, respectively, at momentum $\bk$.
Assuming that initially the system is at zero temperature, the initial energy of each mode is $E_{0\bk}=\omega_{0\bk}/2$;
the equal distribution between the kinetic and potential terms leads to $\left\langle|\Pi_\bk|^2\right\rangle_0=\omega_{0\bk}/4$ and $\left\langle|\phi_{\bk}|^2\right\rangle_0=1/(4\omega_{0\bk})$. Using these values, the expectation value of the energy after the quench is $E_{\bk}=(\omega_{0\bk}+\omega_\bk^2/\omega_{0\bk})/4$. Therefore, the quasiparticle distribution following the quench becomes $n_{\bk}+1/2=E_{\bk}/\omega_{\bk}=(\omega_{0\bk}/\omega_{\bk}+\omega_\bk/\omega_{0\bk})/4$ \footnote{See Refs.~\cite{Gambassi-Mitra15,Mitra-Gambassi16} for a different derivation. The argument presented here follows a private communication with M. Foss-Feig.}.
Clearly, $n_\bk=0$ in the absence of the quantum quench when $\omega_{0\bk}=\omega_\bk$.
For a quench from an initial state {\it far} from the critical point,
$\omega_{0\bk}\approx \omega_0$ is a {\it large} constant
independent of $\bk$, and thus the quasiparticle distribution becomes $n_\bk\approx \omega_0/(4\omega_\bk)$.
Comparing against a thermal distribution at {\it high } temperatures, $n_\bk \approx T/\omega_\bk$, one finds an {\it effective temperature} $T_{\rm eff}\approx \omega_0/4$ \cite{Gambassi-Mitra15,Mitra-Gambassi16}.
Nevertheless, the emergence of an effective temperature does not indicate a fully thermal behavior at short times. In fact, the system may be quenched to a prethermal critical point with the dynamic exponent $z=1$ reminiscent of quantum phase transitions.
Therefore, the dynamics will have mixed features that are partially thermal
and partially quantum
\cite{Chandran13,Mitra-Gambassi15,Gambassi-Mitra15,Mitra-Gambassi16}.
However, to fully expose quantum features, we propose a quench protocol that circumvents the emergence of an effective temperature.

\begin{figure}
  \includegraphics[width=5.5cm]{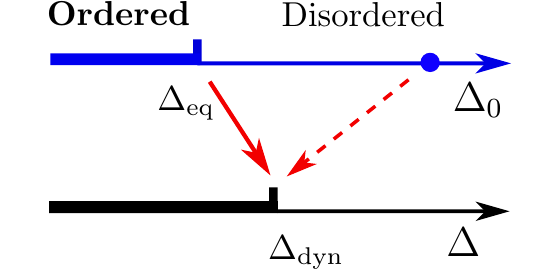}
  \caption{A quantum quench to a prethermal critical point; thick lines represent ordered phases, $\Delta_0$ ($\Delta$) denotes the initial (final) value of the control parameter, and
$\Delta_{\rm eq}$ ($\Delta_{\rm dyn}$) denotes the equilibrium (prethermal or dynamical) critical point.
The dashed arrow indicates a quantum quench from a point far away from $\Delta_{\rm eq}$, and leads to a partial thermalization. The solid arrow indicates the quench $\Delta_{\rm eq}\to\Delta_{\rm dyn}$, and gives rise to a novel dynamical universality class. (The dependence of $\Delta_{\rm dyn}$ on the initial value $\Delta_0$ is neglected in this schematic picture.)}\label{Fig. Quench protocols}
\end{figure}

{\it Quench protocol.---}This work is based on a different quench protocol (solid vs. dashed arrow in Fig.~\ref{Fig. Quench protocols}) where the initial state is at a quantum critical point. In this case, $\omega_{0\bk}$ vanishes at long wavelengths as $\bk \to 0$, and subsequently the post-quench quasiparticle distribution is no longer thermal ($n_\bk\nsim 1/\omega_\bk$); neither does it correspond to zero temperature ($n_\bk\ne0)$.
The fine-tuning of both initial and final parameters is reminiscent of the double fine-tuning (of both temperature and mass) in equilibrium quantum phase transitions \cite{SachdevBook}.
Indeed we show that a {quantum quench from an initial quantum critical point to a prethermal critical point} leads to a dynamical universality class of a genuinely quantum nature.
In particular, we consider a generic situation where the dispersion relation is linear at both pre- and post-quench critical points, $\omega_{0\bk}=c_0 |\bk|$ and $\omega_\bk=c|\bk|$;
here, $c_0$ and $c$ play the role of the ``speed of light'' before and after the quench, respectively. In a sense, in the protocol described here, the speed of light is suddenly quenched at the critical point.
At a microscopic level, this can be achieved, for example, by manipulating the hopping amplitude on a lattice which directly affects the propagation of local propagations.
A generic example with such properties is provided by the $O(N)$ model introduced below.

We should remark that quantum quenches from a gapless to another gapless phase has also been studied in the context of the Luttinger model and its close relatives as well as Fermi liquids \cite{Cazalilla06,Cazalilla09,Cazalilla14,Cazalilla16review,Kehrein08,Kehrein09}.
However, the latter do not fall in the paradigm in Fig.~\ref{Fig. Quench protocols} since the critical behavior (e.g., critical exponents) discussed in these contexts
has been limited to quasi-long-range order in low dimensions. In contrast, the focus of this paper is to study critical points corresponding to truly-ordered phases.

{\it Model.---}Due to their simplicity and generality, we consider the $O(N)$ models which describe, among other things,
the (near-)critical behavior of Ising models, Josephson junction arrays, and quantum antiferromagnets \cite{SachdevBook}.
The Hamiltonian of the $O(N)$ model in $d$ dimensions is given by
\begin{equation}
  H= \int d^d\bx \left[\frac{1}{2}\bpi^2 +\frac{c^2}{2}(\nabla\bphi)^2 + \frac{r}{2}{\bphi}^2 + \frac{u}{4! N}(\bphi^2)^2\right],
\end{equation}
where $\bphi=(\phi_1, \cdots, \phi_N)$ represents an $N$-component scalar field and $\bpi$ denotes its conjugate momentum. Moreover, $u$ is the interaction strength (normalized by $N$), $r$ controls the distance from the critical point, and $c$ denotes the ``speed of light''.
This model has a Lorentz symmetry, and whose dispersion relation is linear, $\omega_\bk= c|\bk|$, at its (equilibrium) critical point. A crucial property of this model is that it is integrable for $N=\infty$ \cite{Sciolla13,Chandran13,Smacchia15}, and thus a long-lived prethermal state emerges for large values of $N$.

\begin{figure}
  \includegraphics[width=5.5cm]{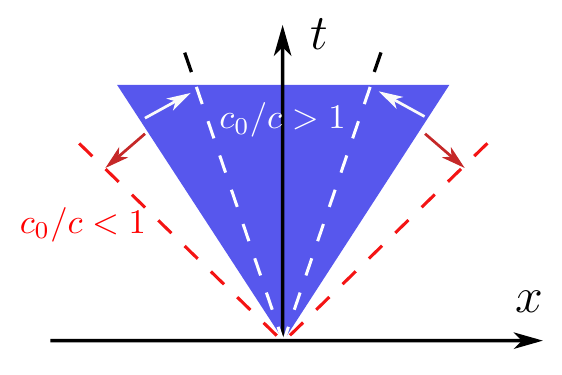}
  \caption{Quantum quench of the ``speed of light'' $c_0\to c$. The causal region before the quench is highlighted, arrows indicate the quench protocol, and the dashed lines represent the post-quench light cones for the two cases $c_0/c>1$ and $c_0/c<1$. The causal region
may shrink ($c_0/c>1$) or expand ($c_0/c<1$) upon the quantum quench. The short-time universal behavior in the two cases is qualitatively different.}\label{Fig. Causal region quenched}
\end{figure}

The quantum quench described earlier is
one that the speed of light suddenly changes from $c_0$ at an initial equilibrium {\it critical} point to $c$ at a prethermal critical state (the latter yet to be determined).
While the critical behavior of many-body systems in equilibrium depend on few features such as dimensionality and symmetry, those describing a prethermal state may also depend on the relative values of pre- and post-quench parameters.  Indeed we shall see shortly that critical exponents depend on the ratio $c_0/c$ in a nontrivial way. Moreover, we find that the dynamics is qualitatively different depending on whether $c_0/c<1$ or $c_0/c>1$ where the causal region defined by the light cone expands or shrinks upon the quantum quench, respectively (see Fig.~\ref{Fig. Causal region quenched}).

{\it Prethermal critical state via RG.---}The nonequilibrium nature of the quench problem necessitates a framework suited for such situations. A powerful tool is the Keldysh functional integral that, together with renormalization group (RG) theory, allows us to systematically study time evolution as well as nonequilibrium stationary states. We first focus on the critical properties of the stationary prethermal state reached at late times. We shall relegate technical details to the Supplemental Material (SM), and report the RG equations at one loop:
\begin{align}
\begin{split}
   \frac{dr}{dl}&=2r+\frac{N+2}{24N} a_d u \Lambda^d \left(\frac{1}{c_0\Lambda}+\frac{c_0 \Lambda}{c^2\Lambda^2+r}\right) +{\cal O}(u^2), \!\!\!\!\!\!\!\!\!\!\!\! \\
   \frac{du}{dl}&=(3-d)u-\frac{N+8}{48N}\,\frac{a_d u^2}{c^3}\left(\frac{c}{c_0}+\frac{c_0}{c}\right)+{\cal O}(u^3),
\end{split}
\end{align}
with $a_d=2/[(4\pi)^{d/2}\Gamma(d/2)]$. The speed $c$ is unrenormalized at one loop, and possibly to all orders, yielding the dynamic exponent $z=1$. The RG equations are written in a standard notation where space and time are rescaled by a factor $b=e^{-l}$ and $\Lambda$ defines the UV cutoff. As is typical in nonequilibrium situations, we have chosen a renormalization scheme where a momentum shell integral $e^{-l}\Lambda < |\bk| \le \Lambda$ is replaced by one that involves a soft cutoff; see the SM for details. Several characteristics of these RG equations are in order:
(i) The upper critical dimension is $d=3$, which will justify a perturbative RG analysis in $\epsilon=3-d$; this is also a feature of equilibrium quantum phase transitions of the $O(N)$ model at zero temperature, therefore it serves as an indication towards a genuinely quantum behavior in our nonequilibrium setting.
(ii) The RG equations (and critical exponents discussed shortly) depend on the initial state via $c_0$.
This is a general feature of prethermal states as they retain some memory of the initial state and are thus in a sense less universal than thermal states in that they depend on more parameters.
Properties (i) and (ii) indicate genuine quantum and dynamical features of the prethermal state.
(iii) To the order considered here, the RG equations describe both an initial state at the Gaussian fixed point $(r_0=0,u_0=0)$ as well as one at the Wilson-Fisher (WF) fixed point of the interacting model $(r_{0\rm WF}, u_{0\rm WF})$ with the subscript $0$ corresponding to $c_0$; see the SM for an explanation.
(iv) To the lowest order in $u$, the RG equations at $c_0=c$ reduce to those in equilibrium (at comparable order using the same renormalization scheme). Although $r$ and $u$ may still take different values before and after the quench, their effect appears only at higher orders in $u$.

To study the RG equations in more detail, we first determine the fixed point of the RG flow.
To the first order in $\epsilon$, we find
\begin{align}
\begin{split}
  r_{\rm dyn}&=-\frac{N+2}{N+8}\,c^2\Lambda^2\,\epsilon +{\cal O}(\epsilon^2),  \\
  u_{\rm dyn}&=\frac{N}{N+8}\,\frac{96\pi^2c^3}{c/c_0+c_0/c}\,\epsilon +{\cal O}(\epsilon^2).
\end{split}
\end{align}
Note that generally $(r_{\rm dyn}, u_{\rm dyn})\ne(r_{\rm WF}, u_{\rm WF})$, the latter being the equilibrium WF fixed point corresponding to $c$ which, to this order, is obtained by setting $c_0\to c$ in the above equation. Therefore, $(r_{\rm dyn}, u_{\rm dyn})$ describes a new, dynamical fixed point.
The RG equations can be linearized around the fixed point to find the corresponding eigenvalues, which in turn determine the critical exponents. We find that the exponent $\nu$ characterizing the divergence of the correlation length with the distance from the critical point ($\xi \sim \delta r^{-\nu}$) is given by
\begin{equation}
 \nu
 =\frac{1}{2} +\frac{1}{2}\frac{N+2}{N+8}\frac{1}{c^2/c_0^{2}+1}\,\epsilon + {\cal O}(\epsilon^2).
\end{equation}
Remarkably, this exponent depends (continuously) on the ratio $c_0/c$, hence the universal dependence on the initial state.
\begin{figure}
  \includegraphics[width=7.5cm]{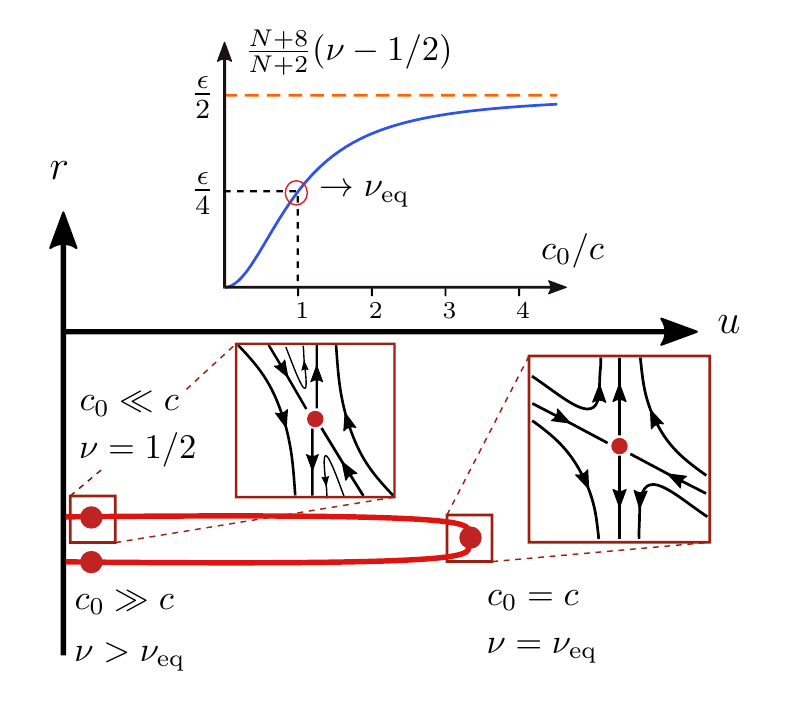}
  \caption{Fixed points and RG flow of the prethermal phase reached upon a quench of the speed of light from $c_0$ to $c$. The thick (red) curve schematically plots the one-parameter family of fixed points $(r_{\rm dyn},u_{\rm dyn})$ as a function of $c_0/c$ computed to the first order in $\epsilon=3-d$; while $r_{\rm dyn}$ is constant in $c_0/c$ to this order, a slight dependence is assumed for the sake of illustration. The extreme limits (red solid circles), the RG flow in their vicinity (inside the square boxes), and their corresponding exponent $\nu$ are highlighted.
  The dependence of the exponent $\nu$ is plotted in the top panel. For $c_0=c$,
  this exponent reduces to its equilibrium value $\nu_{\rm eq}$.  }
  \label{Fig. RG}
\end{figure}
Also, for $c=c_0$, it assumes its equilibrium value $\nu_{\rm eq}$ to the leading order. Finally, the exponent $\nu$ approaches its mean-field value $\nu=1/2$ to the first order in $\epsilon$ as $c_0/c \to 0$. To see why this is so, we first note that the exponent $\nu$ is related to the renormalization of the mass term $r$. To the lowest order, the latter is renormalized by $\Delta r\sim u \langle \bphi^2\rangle$. For small $c_0$, the field $\bphi$  is highly fluctuating before the quench, and $\langle \bphi^2 \rangle$ is dominated by the contribution due to its pre-quench value, and thus would be less sensitive to the mass term $r$ in the final Hamiltonian. The latter implies that the renormalization correction can be simply absorbed in the fixed point value of $r$, therefore not changing  $\nu$ away from its mean-field value.

The RG equations, their fixed point, and the corresponding eigenvalues are reminiscent of the WF fixed point of a quantum critical point with the dynamic exponent $z=1$; however, the emergence of the ratio $c_0/c$ defines a new {\it quantum dynamical universality class} with distinct properties. In Fig.~\ref{Fig. RG}, we have shown the one-parameter family of fixed points as a function of $c_0/c$ and the corresponding RG flow at some, including extreme, values of $c_0/c$; the exponent $\nu$ is also plotted as a function $c_0/c$ to the first order in $\epsilon$. The topology of the RG flow is the same as that of the WF fixed point. In particular, there is only one relevant direction away from any fixed point, and therefore, to access the prethermal critical state, only one parameter ($r$) should be tuned at the quench. (The other fine-tuning is due to the choice of the initial state at the critical point.) However, the corresponding eigenvectors and eigenvalues of the RG flow are different and depend on $c_0/c$.

Finally, we stress again that a long-lived prethermal state emerges for large values of $N$, while dissipative terms appear over a time scale that diverges with $N$ \cite{Sciolla13,Chandran13,Smacchia15,Mitra-Gambassi15,Mitra-Gambassi16,Gambassi-Mitra15,Mitra11,Mitra12}. Therefore, the equations in this paper should be properly understood as a series expansion in $1/N$. As is typical in a number of physical contexts, it may be that even a fairly small value of $N$---such as $N=2,3$ readily accessible in experiments---is large enough to exhibit the features reported here.

\begin{figure}
  \includegraphics[width=7.5cm]{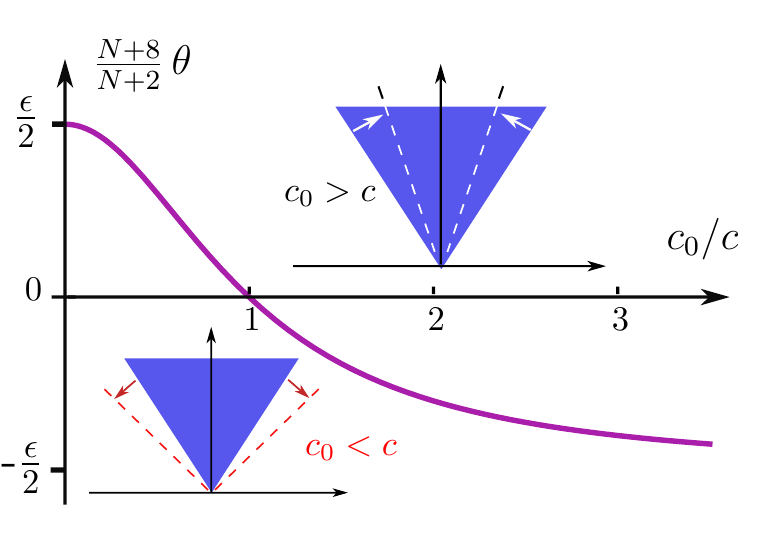}
  \caption{The exponent $\theta$ characterizing the early-time universal behavior after the quench as a function of $c_0/c$. This exponent characterizes the dependence of the response to a perturbation at early times. $\theta$ changes sign at $c_0=c$: For $c_0<c$ ($c_0>c$) where the causal region expands (shrinks), it is positive (negative) indicating a pronounced (suppressed) response to an early-time perturbation; see the text for the explanation.}
  \label{Fig. theta}
\end{figure}

{\it Short-time universal behavior.---}Heretofore, we have studied the system as it approaches a quasi-stationary prethermal state.
Nevertheless, one can also probe universal properties pertaining to early dynamics, for example, by inspecting the response of the system to a perturbation at early times; this information can be obtained from the response function $\bG_R(t,t', \bx-\bx')=-i\Theta(t-t')\langle [\bphi(\bx,t),\bphi(\bx',t')]_-\rangle$. At later times when the system has reached a prethermal critical state, the response function only depends on the difference between times $t$ and $t'$, and its Fourier transform in momentum space ($k\equiv |\bk|$) takes the scaling form $\bG_R(t,t',\bk)=k^{-2+\eta+z}{\bm f}\left(k^z(t-t') \right)$ with $\bm f$ a scaling function, $\eta$ the anomalous scaling dimension, and $z$ the dynamic exponent; in the prethermal state discussed here, $\eta={\cal O}(\epsilon^2)$ and $z=1$.
On the other hand, the response function to a perturbation at early times should explicitly depend on both $t$ and $t'$. In the limit $t'/t \to 0$, the response function to the leading order
takes the scaling form $\bG_R(t \gg t',\bk)=(t/t')^\theta k^{-2+\eta+z}{\bm g}(k^z t)$ where a new exponent $\theta$ emerges along with a scaling function $\bm g$ characterizing the corresponding universal behavior \cite{Schmalian14,Schmalian15,Gambassi-Mitra15,Mitra-Gambassi16}.
We find the exponent $\theta$ from a perturbative one-loop RG calculation as
\begin{equation}
  \theta=\frac{1}{2}\frac{N+2}{N+8}\,\frac{c/c_0-c_0/c}{c/c_0+c_0/c}\,\epsilon + {\cal O}(\epsilon^2).
\end{equation}
This equation describes a genuinely quantum short-time exponent with no analog in either classical or, past studies of, quantum systems. Given that the short-time behavior is tied to the critical properties of the long-lived prethermal state, the exponent $\theta$, similar to those at the prethermal critical state, also depends continuously on the ratio $c_0/c$. Interestingly, this exponent has opposite signs for the two regimes $c_0/c<1$ and $c_0/c>1$, see Fig.~\ref{Fig. theta}.
The reason for such behavior can be understood on general grounds. First note that the response function $\bG_R(t,t')$ (momentum dependence is suppressed) characterizes the response of the system to an infinitesimal magnetic field at the initial time $t'$.
This function may be expected to fall off in time; however, the size of the causal region also changes after the quench.
Indeed, for $c_0<c$, the causal region expands after the quench, which tends to amplify the response function. This is consistent with the fact that $\theta>0$ for $c_0/c<1$ resulting in an increase of $\bG_R (t\gg t') \sim (t/t')^\theta$ with  $t/t'$. In the opposite regime where $c_0/c>1$, the causal region shrinks after the quench that leads to a decrease of the response function with $t/t'$.
In a similar way, one can also characterize the dependence of the correlation function $\bG_K(t,t', \bx-\bx')=-i\left\langle [\bphi(\bx,t),\bphi(\bx',t')]_+\right\rangle$. It turns out that the correlation function $\bG_K(t\gg t')\sim (t/t')^\theta$ also has the same universal dependence on $t/t'$, at least to the lowest order in $\epsilon$ expansion. The dependence on short times, and the fact that the same exponents arise in both response and correlation functions is different from classical aging \cite{Janssen89} as well as quantum aging in a deep quench from a noncritical initial state \cite{Mitra-Gambassi15,Gambassi-Mitra15,Mitra-Gambassi16}, but is similar to the short-time universal behavior in a system coupled to a zero-temperature bath \cite{Schmalian14,Schmalian15}.

{\it Conclusions and Outlook.---}In this work, we have considered quench dynamics in a quantum system, and have studied the long-lived prethermal state emerging before the onset of thermalization at long times. Crucially, we have considered a quench protocol where an initial critical state is suddenly quenched to a prethermal critical state with the double fine-tuning reminiscent of quantum phase transitions. We have shown that the prethermal critical state falls under a genuinely quantum dynamical universality class, and furthermore the dynamics at short times follows a distinct universal behavior that mimics a kind of quantum aging.
While we have focused on the paradigmatic $O(N)$ model,
our general conclusions should apply to a generic situation where the system approaches a long-lived prethermal state. Nearly integrable models, including those in 1D, would be interesting candidates to further probe the universal dynamics at short times after a quantum quench.

The author acknowledges start-up funding from Michigan State University.

\newpage
\null
\newpage

\setcounter{equation}{0}
\renewcommand{\theequation}{S\arabic{equation}}
\setcounter{figure}{0}
\renewcommand{\thefigure}{S\arabic{figure}}

\onecolumngrid
\begin{center}
\Large\textbf{Supplemental Material}
\end{center}

\twocolumngrid

In this supplement, we derive the RG equations describing the prethermal state (Sec.\ \ref{App: RG}), and also derive the short-time exponent $\theta$ and the scaling form of the response function (Sec.\ \ref{App: theta}).

\section{RG equations}\label{App: RG}
We start by setting up the Hamiltonian dynamics within the framework of the Keldysh formalism. The Keldysh action on the closed-time contour reads as $S=\int_t  (L[\bphi_+]-L[\bphi_-])$ where the subscripts $\pm$ denote the forward/backward branches of the contour, and $L[\bphi]=\frac{1}{2}\int_\bx \big[ \dot\bphi^2-c^2(\nabla \bphi)^2-r\bphi^2-(2u/4N!)(\bphi^2)^2\big]$ is the Lagrangian corresponding to the Hamiltonian in Eq.~(1) of the manuscript. It is more convenient to write the action in the Keldysh basis defined as $\bphi_{c/q}=(\bphi_+\pm \bphi_-)/\sqrt{2}$.
The action then reads
\begin{align}
  S=&\int_0^\infty \!\!\!\!dt \int d^d\bx\Big[\dot\bphi_q\cdot \dot \bphi_{c} -c^2\nabla\bphi_q\cdot\nabla\bphi_q-r \bphi_q\cdot \bphi_{c} \nonumber \\
  &-\frac{2u}{4!N}(\bphi_q\cdot \bphi_{c}) \bphi_{c}^2-\frac{2u}{4!N}(\bphi_q\cdot \bphi_{c}) \bphi_{q}^2 \Big].
\end{align}
In principle the coefficients of the two interaction terms may evolve differently under RG.
However, we shall see that, at least to the order considered here, their RG flow is indeed the same, which is reminiscent of a quantum critical point. The response and correlation functions can be represented in the Keldysh basis as $i \bG_{R/K}(t,t',\bx-\bx')=\langle \bphi_c (t,\bx)\bphi_{q/c}(t',\bx')\rangle$. Due to the $O(N)$ symmetry, cross correlations are zero, and therefore we consider the propagators $G_{R/K}$ corresponding to a single component of the field $\bphi$. We further adopt the notation $g_{R/K}$ for the response and correlation functions at the quadratic level of the action; the latter are given by \cite{Mitra-Gambassi15}
\begin{align}
\begin{split}
  &g_{R}(t-t',\bk)= -\Theta(t-t')\frac{\sin\omega_\bk(t-t')}{\omega_\bk}, \\
  &ig_{K}(t,t',\bk)= \frac{K_+\cos(\omega_\bk(t-t'))+K_-\cos(\omega_\bk(t+t'))}{\omega_\bk},
\end{split}
\end{align}
with $K_{\pm}=\left({\omega_\bk}/{\omega_{0\bk}}\pm{\omega_{0\bk}}/{\omega_\bk}\right)/2$.
Note that the response function at the quadratic level, $g_R$, only depends on the time difference, $t-t'$. To find the scaling dimensions at the critical point where $\omega_\bk=c|\bk|$, we first note that the equal-time correlation function inside the light cone, $|\bx-\bx'|\lesssim 2ct$, approaches a stationary value given by
\begin{align}
  ig_{K}\left(t,t,\bx-\bx'\right) &\sim \int d^d \bk \frac{K_+}{\omega_\bk} e^{i\bk\cdot (\bx-\bx')}
  \sim \frac{\bar K_+ }{|\bx-\bx'|^{d-1}},
\end{align}
where the constants $\bar K_\pm\equiv (c/c_0\pm c_0/c)/2$ are defined as the value of $K_{\pm}$ at the critical point. Together with the quadratic part of the Keldysh action, this equation determines the scaling dimensions of the fields at the quadratic level as
\begin{equation}
  [\bphi_c]=[\bphi_q]=\frac{d-1}{2},
\end{equation}
resembling those of a quantum phase transition in $d$ dimensions with $z=1$.
Using these scaling dimensions, the coefficients $r$ and $u$ scale as $r\to b^2 r$ and $u\to b^{3-d} u$ under rescaling space and time coordinates.
Next we consider the renormalization of the coefficients $r$ and $u$ by integrating out higher-momentum modes. However, one should choose a smooth cutoff to avoid nonsensical results. At the one-loop level, we choose a simple scheme where the conventional momentum-shell integral over $|\bk| \in (\Lambda-\Delta \Lambda, \Lambda]$ is substituted by $\int_> (\cdot)\equiv\int_\bk \left[h(|\bk|/\Lambda)- h(|\bk|/(\Lambda-\Delta\Lambda))\right] (\cdot)$ where the dot represents the (momentum-dependent) integrand, and $h(x)$ is a smooth function such that $\lim_{x\to0}h(x)=1$ while it decays sufficiently rapidly as $x\to\infty$; for example, $h_n(x)=e^{-x^n}$ for a positive $n$.
Using this scheme, we obtain the renormalization of the mass term $r$ to the first order in $u$ as
\begin{align}
  \Delta r&= \frac{N+2}{12N}u \int_{\!>} ig_{K} (t,t,\bk)\nonumber \\
  &= \frac{N+2}{12N}\frac{ua_d \Delta \Lambda}{c\Lambda}\Big[\bar K_+ \Lambda^{d-1}\Gamma\big(\frac{d+n-1}{n}\big)\nonumber \\
       &\hskip 1in -\frac{c_0}{2c^3}\,r+\cdots\Big].
\end{align}
We have computed the integral to the appropriate order in $\epsilon$ since we are interested in the vicinity of the fixed-point values of $r$ and $u$ which are both anticipated to be of the order ${\cal O}(\epsilon)$. The dots include terms of higher orders in $\epsilon$ and those that decay with time.
Together with the scaling of $r$, the above equation yields the RG equation for $r$. In the limit of $n\to \infty$ and up to the order of ${\cal O}(\epsilon^2)$, this is consistent with the first equation in Eq.~(2) of the manuscript, which is simply computed by imposing a hard cutoff. Different values of $n$ will change the fixed-point value of $r$ but do not affect its universal properties and the corresponding critical exponents such as $\nu$.

A nontrivial renormalization of $u$ starts at the second order in $u$,
\begin{align}
  \Delta u &=\frac{N+8}{6N}u^2\int_0^\infty \!\!\!\!ds \int_{\!>} \left[ig_{K}(t,s,\bk) g_{R}(t-s,\bk)\right].
\end{align}
A simple calculation shows that this quantity approaches a constant at long times ($\Lambda ct\gg1$). We find, to the leading order in $\epsilon$ and independent of the form of the cutoff function $h$,
\begin{equation}
  \Delta u = -\frac{N+8}{24N}\frac{a_d u^2\Delta \Lambda}{ c^3 \Lambda} \bar K_+.
\end{equation}
Together with the scaling behavior of $u$, this equation leads to the RG flow in the second line of Eq.~(2) of the manuscript.

While the above analysis was done for an initial state at the Gaussian fixed point ($r_0=u_0=0$), the resulting RG equations would be the same at the Wilson-Fisher fixed point of the initial state to the order reported. This is because any such correction would involve the difference $G_{0K}- g_{0K}$ where $G_{0K}$ and $g_{0K}$ denote the equal-time correlation functions of the initial state (hence the subscript $0$) with and without interactions, respectively. However, at the critical point of the initial state, this difference only arises at the order ${\cal O}(\epsilon^2)$ because the exponent $\eta\sim {\cal O}(\epsilon^2)$. Therefore, the fixed-point values as well as the exponents do not change to the first order in $\epsilon$.

\section{Short-time exponent}\label{App: theta}
Suppose that the quench is tuned to bring the system to its (prethermal) critical point at long times, i.e., $r(t) \to r_{\rm dyn}$ as $t\to \infty$. The distance from the critical point at time $t$, defined as $\delta r(t)\equiv r(t)-r_{\rm dyn}$, is given to the first order in $\epsilon$ by
\begin{equation}
  \delta r(t)= \frac{N+2}{12N}u \int_\bk \left[ig_{K} (t,t,\bk)-ig^{\rm dyn}_{K}(\bk)\right],
\end{equation}
where $g^{\rm dyn}$ denotes the late-time limit of $g$ evaluated at the prethermal critical state.
This integral should be evaluated using the soft cutoff introduced above. We find, to the leading order in $\epsilon$ and independent of the form of the cutoff function $h$,
\begin{equation}
  \delta r(t) =-\frac{N+2}{12N}\frac{u a_3 \bar K_-}{4c^3t^2}\equiv -\frac{\theta}{t^2},
\end{equation}
where we have defined the dimensionless constant $\theta$ to be identified with the short-time exponent. At the (dynamical) critical point, we find $\theta=\frac{N+2}{N+8}(\bar K_-/\bar K_+)\epsilon/2$ from which Eq.~(5) of the manuscript follows. We still have to show that the response function scales with time as $(t/t')^\theta$ for $t\gg t'$. To this end, we compute the retarded Green's function to the next order as
\begin{align}
  G&_{R}(t,t', \bk)=g_{R}(t-t', \bk) \nonumber \\
  &+\int_0^\infty \!\!\!\!ds \left[ g_{R}(t-s, \bk) \delta r(s) g_{R}(s-t', \bk)\right]+\cdots.
\end{align}
The integral in the second line is particularly simple in the limit $\bk \to 0$ where $g_{R}(t-t',\bk\to 0)=-(t-t')$ for $t>t'$. Using the above equations, we find
\begin{equation}
  G_R(t\gg t',\bk\to0)\approx -t\left[1+\theta \log(t/t')\right].
\end{equation}
The logarithm in this equation can be exponentiated to produce the scaling behavior reported in the manuscript. A similar argument follows for the correlation function $G_K(t\gg t',\bk\to 0)$.

\end{document}